# Manufacturable 300mm platform solution for Field-Free Switching SOT-MRAM

K. Garello, F. Yasin, H. Hody, S. Couet, L. Souriau, S. H. Sharifi, J. Swerts, R. Carpenter, S. Rao, W. Kim, J. Wu, K. K. V. Sethu, M. Pak, N. Jossart, D. Crotti, A. Furnémont, G. S. Kar

imec, 3001 Leuven, Belgium; email: kevin.garello@imec.be

**Abstract** — We propose a field-free switching SOT-MRAM concept that is integration friendly and allows for separate optimization of the field component and SOT/MTJ stack properties. We demonstrate it on a 300 mm wafer, using CMOS-compatible processes, and we show that device performances are similar to our standard SOT-MTJ cells: reliable sub-ns switching with low writing power across the 300mm wafer. Our concept/design opens a new area for MRAM (SOT, STT and VCMA) technology development.

**Introduction:** Among non-volatile memory technologies, Spin-Transfer-Torque (STT) MRAM is seen as a credible candidate to replace SRAM in low level caches due to its scalability, low power and high speed, as well as compatibility with scaled CMOS processes and voltages. This is reflected by major foundries and tool suppliers investing significant R&D resources into embedded MRAM past years. Recently they even started prototyping demonstrators, progressively reaching maturity for mass production [1-5]. However, STT-MRAM cannot reliably operate at sub-ns scales due to large incubation delays. It can suffer from reliability issues due to the shared read/write terminal and large write voltage stress, making it an unsuitable solution to tackle L1/2 SRAM cache replacement and non-volatile logic. In the search of faster writing mechanisms, spin-orbit torque (SOT) [6-8] is seen as a promising solution. SOT induces switching of the free layer (FL) of the MTJ by injecting an in-plane current in an adjacent SOT layer, typically with the assistance of a static in-plane magnetic field to insure deterministic magnetization switching. The need for this external field is a major hurdle for practical MRAM products. To circumvent it, various solutions for "Field-Free Switching" (FFS) have been proposed, e.g., different shapes [9], in-plane MTJ [10], and anti-ferromagnets [12]. However, all these solutions compromise scalability and/or SOT efficiency. Here, we propose and demonstrate an integration-friendly approach to FFS via embedding a ferromagnet in the hardmask that is used to shape the SOT track. This magnet can be realized as BEOL metal line [13]. Our proposal allows to separately optimize the SOT material, MTJ properties, and FFS conditions. Our results show that FFS-SOT-MTJ cells keep its original performances, namely sub-ns reliable switching with low writing power, opening a new area for SOT-MRAM technology development.

**Concept:** In Fig.1, we show our FFS design concept of FFS-SOT cell and a representative TEM. It is a SOT-MTJ device, where we use a magnetic hard mask (MHM) to shape the SOT track. The rectangular shape of MHM provides strong shape anisotropy along the current direction, therefore generating a maximized and homogeneous in-plane field $B_{MHM}$ on the free-layer (FL). The integration process is very similar to our previous report [7], with the novelty of adding access VIAS between bottom electrode and SOT track, and adding a process step during SOT hard mask module to deposit a 50nm thick Co layer prior to the traditional TiN hardmask, forming a MHM. In Fig. 2.a, using micromagnetic simulations we quantify $B_{MHM}$ for a 50 nm thick Co MHM as a function of height Z with respect to the FL position. This accounts for process variation of MTJ hardmask consumption. For our standard process (Z~80nm), we estimate the field on FL to be $|B_{MHM}|=36$ mT, which is sufficient for achieving FFS. As shown in Fig. 2.b, $B_{MHM}$ is very homogeneous at FL location. The MTJ used in this study is a top-pinned stack SOT/CoFeB(1nm)/MgO(RA50)/CoFeB/SAF annealed at 300°C, that is perpendicularly magnetized (PMA), and where the SOT layer is 3.5nm of Tungsten. In the following, the typical structure studied consist of a 60 nm circular MTJ sitting on a 170nm wide SOT track. VIA gap is 140nm, resulting in a SOT resistance of ~270 Ω ($\rho_{SOT}$~160 μΩ.cm), and MHM is $120*390$ nm$^2$. We use dense dummification, with a pitch of $260*540$ nm$^2$ to study the neighboring magnetic field impact.

**FFS demonstration**: In Fig. 3, we first show standard R-V curves under large applied in-plane field $|B_x|=50$mT. We obtain classical switching conditions for negative spin hall angle of W (Fig. 3.a): a positive voltage $V_{SOT}$ switches the MTJ state from parallel (P) to anti-parallel (AP) state at $B_x=+50$mT (black curve), while $B_x=-50$mT results in AP→P transition (red curve). Secondly, we present R-V curves under zero applied field (Fig. 3.b) and show that switching is also reliably obtained. Prior to the zero-field measurement, the MHM is saturated with a positive (negative) external field. In manufacturing, the MHM field alignment can easily be done at the fab-out. Note that the switching symmetry is now inverted: a positive $V_{SOT}$ switches AP→P (P→AP) for positively (negatively) initialized MHM (black (red) curve). This is because $B_{MHM}$ is opposed to Co MHM field direction $B_{Co}$. The functionality was tested for a large set of devices, ranging from 50 to 100nm; the behavior is maintained all across the three tested wafers, showing the robustness and simplicity of our concept. In Fig. 4, we evaluate $B_{MHM}$ ~ -32mT by measuring the $P_{sw}$ versus $B_x$. It is close to our simulation results.

**Device performances**: In the following, we set $B_{MHM} < 0$, $B_x=B_z=0$. In Fig. 5, we present the R-H curve for a 60nm device, with typical characteristics: 83mT coercivity $B_c$, -11.5mT offset field $B_{off}$, 270mT RL pinning field $B_{pin}$. We estimate from $B_c$ distribution a retention $\Delta = 48$ and an anisotropy $B_k = 190$mT. The stack properties are weakly impacted by MHM compared to our reference stack (table 1). Fig. 6 and 7 shows that FFS devices perform extremely well down to 300ps. At $P_{sw}=50\%$ for a 500ps short pulse, we report for AP-P/P-AP: $V_{SOT}= 0.34/-0.3$V, corresponding to a current of 1.28/-1.1 mA ($J_{sw}=183/-159$ MA/cm$^2$). In Fig. 8, we test 1ns WER down to $10^5$ events, endurance up to $10^{11}$ cycle at 1ns with $V_{SOT}=0.4$V at a repetition rate of 20 MHz. Neither SOT nor MTJ resistances vary, and $P_{sw}$ is only weakly affected post endurance test (20mV shift), evidencing the robustness of our devices under high stress.

**Conclusion**: We propose a novel integration approach that allows to reliably write SOT-MTJ cells without the assistance of an external magnetic field. Our design is based on a magnetic layer embedded in the SOT hard mask. The approach is reliable and integration-friendly and not compromising the original sub-ns writing and low energy performances. That makes SOT-MRAM a viable candidate for the replacement of SRAM at low level cache. Further research will focus on reducing writing current via layout optimization, as well as SOT, stack and device engineering.

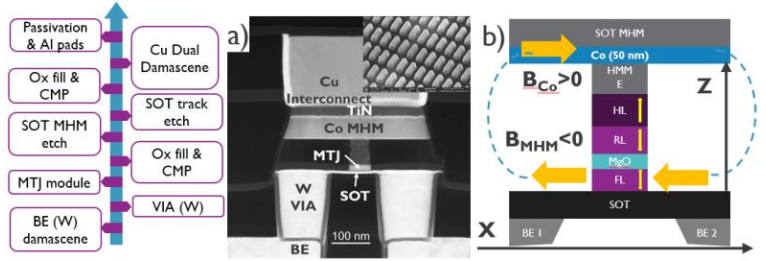

**Fig. 1:** Main integration modules, a) lateral TEM cross section view of the SOT-FFS-MTJ with 50nm thick Co magnetic hard mask (inset is top SEM view after MHM etch), b) schematic of SOT-FFS-MTJ stack with top-pinned SAF design: W(SOT) / CoFeB(FL) / MgO / CoFeB(RL) / SAF (HL).

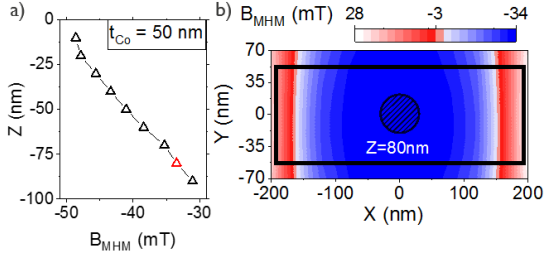

**Fig. 2:** Micromagnetic simulations of $B_{MHM}$ for a 5*5 MHM array (120*390nm$^2$ with pitch of 260*540 nm$^2$) is plotted for the central array track: a) as a function of height Z (red point is typical process), b) [X,Y] position at Z=-80nm. Dot/rectangle represent respectively the FL (60nm) and the SOT track.

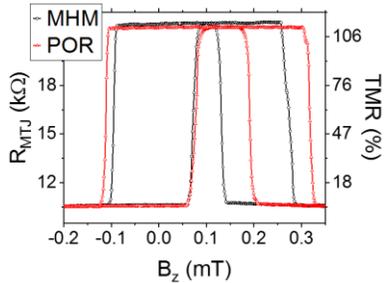

**Fig. 3:** R-V switching curves a) under large applied in-plane field $|B_x|$=50mT, b) **At 0 external field**, after saturating MHM with a positive (negative) external. Switching symmetry is inverted because $B_{MHM}$ is opposed to $B_{Co}$ MHM magnetization direction.

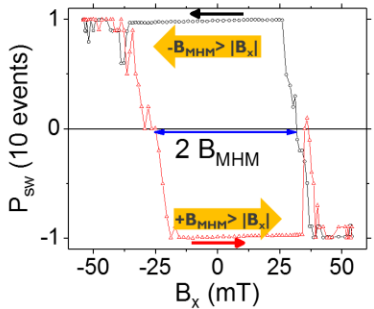

**Fig. 4:** Switching probability $P_{sw}$ vs. in-plane field $B_x$ at $\tau_p$=10ns for 60nm device. There are two zones where $B_x$ and $B_{MHM}$ are competing, resulting in a low $P_{sw}$. Average of the two field regions gives $B_{MHM}$~ 32mT, similar to simulations.

**Fig. 5:** R-H loop as a function of external Z field $B_z$ for SOT-MHM 60nm device and an exact same reference stack (POR) without MHM. $B_{MHM}$ has low impact on FL coercivity and RL pinning field.

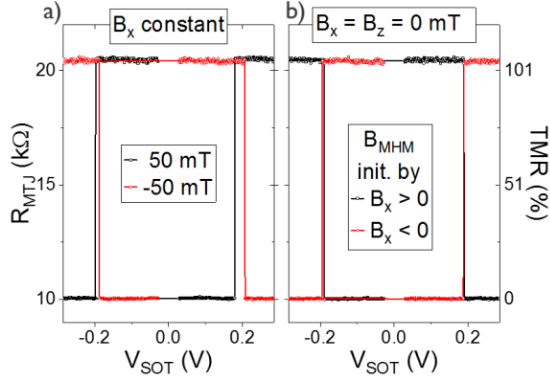

**Fig. 6:** 60nm device at 0 external field. a) $P_{sw}$ as a function of applied voltage measured at 0 external field ($B_{MHM}$<0) for various pulse width $\tau_p$. b) 50% switching current distribution for $\tau_p$=1ns measured randomly across all wafer for more than 130 devices.

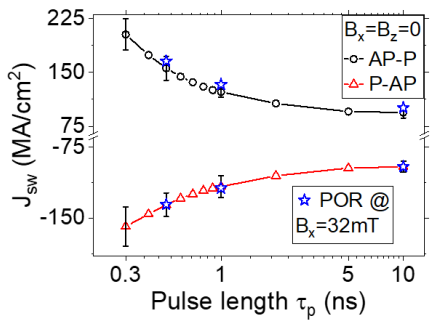

**Fig. 7:** $J_c^{50\%}$ as a function of pulse length $\tau_p$ for 60nm MTJ. Error bars correspond to standard deviation measured for 30 different devices all across the 300mm wafer. Star data represent $J_c^{50\%}$ of the reference stack (POR) without MHM solution, measured for a field $B_x$=32 mT.

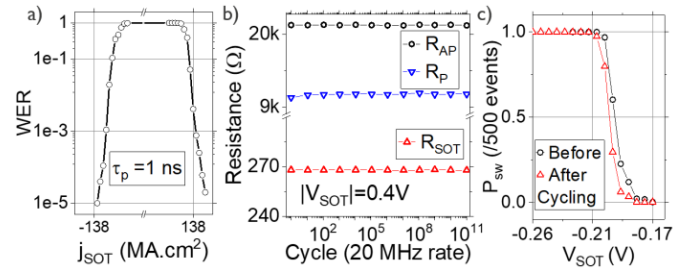

**Fig. 8:** 60nm MHM-MTJ (0 external field). a) Write error rate measured for 10$^5$ events for P-AP and AP-P transitions at 1ns, b) Endurance measured for 10$^{11}$ cycles at a repetition rate of 20 MHz with following conditions: $\tau_p$ = 1ns and $|V_p|$ = 0.4V. No SOT/TMR resitance degradation is observed. c) $P_{sw}$ control measure at $\tau_p$=10ns before and after endurance cycling, showing that device behavior is weakly affected by large cycling (20 mV shift at 50% after cycling).

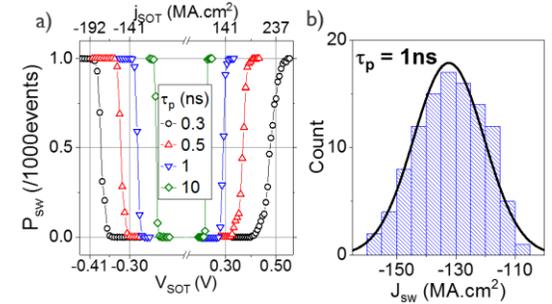

| | POR | MHM |
|---|---|---|
| CD (nm) | 60 | 60 |
| TMR (%) | 110 | 110 |
| RA (Ω.μm$^2$) | 40 | 40 |
| $B_c$ / $B_{off}$ (mT) | 94.5/-16.3 | 83.5/-11.5 |
| $B_x$ (mT) | 320 | 0 |
| Δ / $B_k$ (mT) | 51/200 | 48/190 |
| $\rho_{SOT}$ (μΩ.cm$^2$) | 160 | 160 |
| $|J_{50\%}|_{avg}$ 1 ns (MA.cm$^2$) | 120 | 126 |
| $I_{sw}$ (uA) @1ns & 50nm SOT | 276 | 290 |

**Table 1:** W-based SOT-MTJ cell properties comparison between process of reference (POR) and MHM solutions